# Exploring the Effect of Resolution on the Usability of Locimetric Authentication

Antonios Saravanos[1][0000-0002-6745-810X], Dongnanzi Zheng[1], Stavros Zervoudakis[1], Donatella Delfino[1]

[1] New York University, New York, NY, 10003, USA
{saravanos, dz40, zervoudakis, dd61}@nyu.edu

**Abstract.** Locimetric authentication is a form of graphical authentication in which users validate their identity by selecting predetermined points on a predetermined image. Its primary advantage over the ubiquitous text-based approach stems from users' superior ability to remember visual information over textual information, coupled with the authentication process being transformed to one requiring recognition (instead of recall). Ideally, these differentiations enable users to create more complex passwords, which theoretically are more secure. Yet locimetric authentication has one significant weakness: hot-spots. This term refers to areas of an image that users gravitate towards, and which consequently have a higher probability of being selected. Although many strategies have been proposed to counter the hot-spot problem, one area that has received little attention is that of resolution. The hypothesis here is that high-resolution images would afford the user a larger password space, and consequently any hot-spots would dissipate. We employ an experimental approach, where users generate a series of locimetric passwords on either low- or high-resolution images. Our research reveals the presence of hot-spots even in high-resolution images, albeit at a lower level than that exhibited with low-resolution images. We conclude by reinforcing that other techniques – such as existing or new software controls or training – need to be utilized to mitigate the emergence of hot-spots with the locimetric scheme.

**Keywords:** locimetric authentication, hot-spots, high-resolution graphical passwords.

## 1 Introduction

Locimetric authentication (also known as click-based authentication) is a graphical mechanism that verifies users' identity through their selection of a series of predetermined points on an image in a particular order. Initially described by Blonder [4] in his patent filing (US5559961A), it serves as the first form of graphical authentication. Over the years, several other implementations of the scheme have been developed, such as PassPoints [31], Cued Click Points [6], and Persuasive Cued Click-Points [7]. However, none of these implementations enjoy the level of diffusion as Microsoft's Picture

Password, which is installed by default on any machine running the Windows 8 operating system or higher. In actuality, Picture Password is a combination of two schemes, locimetric and drawmetric, with the user empowered to select how much of each method they prefer to use. Thus, the created password could be fully locimetric, fully drawmetric, or a combination of both schemes. Drawmetric authentication is a form of graphical authentication that validates users by requiring them "to draw a preset outline figure, either on top of an image or on a grid" [10]. Given the Windows operating system's prominence, especially in the desktop market, insight into the potential weaknesses inherent with locimetric authentication is valuable.

In this paper, we focus on one of the accepted weaknesses: the users' propensity to select the same point on images to form their passwords, known colloquially as hot-spots [6] (also sometimes known as click-point clustering [25]). Strategies have been proposed to counter the hot-spot problem by modifying the scheme (see Cued Click-Points [6] and Persuasive Cued Click-Points [8]). However, one area that has not been extensively studied pertains to the resolution of the image that is used [13], the postulation being that high-resolution images would present the user with a larger password space, and therefore any hot-spots would dissipate. Although higher-resolution images have been explored in the past within the context of graphical authentication (e.g., gaze-based authentication [5]), locimetric authentication per-se has received little attention [13]. Taking an experimental approach, we seek to establish whether clustering persists with high-resolution images for locimetric authentication.

## 2 Background

The existence of hot-spots was initially speculated by Wiedenbeck et al. [31], who wrote, "logically, it seems that many users may be attracted to incongruous or unexpected elements in an image". Indeed, while theoretically locimetric authentication has the potential to be superior to text-based authentication [31], if users only select from specific regions, the effectiveness of the scheme drops. Several authors have reported the presence of hot-spots when studying the usability of locimetric authentication [29]. Others have attempted to evaluate whether the weakness can be exploited [11]. The first study we could find examining the usability of locimetric authentication was that by Wiedenbeck et al. [29], who investigated using the ClickPoints implementation while relying on images with a resolution of 451 by 331 pixels. When their study [29] was conducted in 2005, this resolution could be described as adequate. At the time of writing, it is considered a particularly low resolution. To account for backward compatibility, later studies retained the low-resolution specification. This includes other evaluations using PassPoints [30], web-based simulations inspired by PassPoints [26], Java-based simulations inspired by PassPoints [11], and Persuasive Cued Click-Points [8].

Dirik et al. [11] acknowledge the importance that resolution plays in the suitability of an image for use with locimetric authentication, writing that there are "a vast number of possibilities, if the image is large and complex, and if it has good resolution". Accordingly, increasing the resolution should then resolve the hot-spot problem. Indeed, as the resolution increases, there would hypothetically be more potential points for users to click on for their password. We were able to find one paper that examines high-resolution images within the context of the Picture Password mechanism: Gao et al. [13], who undertook a holistic evaluation of the usability of Microsoft's Picture Password. Simulating the Windows 8 operating system, the authors do not explicitly state the size of the images used, although they do disclose that their experiment was conducted on "a PC with a 19-inch screen and 1024 x 1280 screen resolution". The authors do report the presence of hot-spots in all three of the images that they studied.

## 3 Methodology

We employed an experimental approach to evaluate the effect that usability plays on the security of locimetric authentication, looking at two image resolutions: low-resolution (451 by 331 pixels) and high-resolution (1280 by 720 pixels). A series of web-based experiments were held in which participants were each asked to generate seven locimetric passwords based upon preselected images, using software designed to simulate the password setup phase (see Figure 1). The images selected were of a complex composition in order to ensure that participants had adequate points throughout each image to choose, thereby addressing the concern raised by Dirik et al. [11] vis-à-vis image suitability. Each password was comprised of five points; following its creation, participants were asked to reinput their password for verification. The number five was selected as the appropriate password length based on the majority of existing research examining the usability of locimetric authentication, which uses as a base Wiedenbeck et al.'s [31] PassPoints.

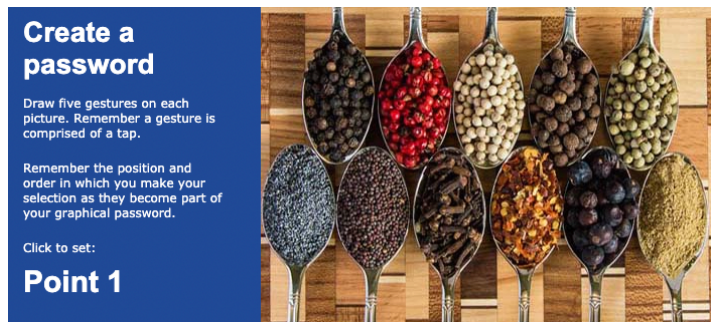

**Fig. 1.** Screen capture of software designed to simulate the creation of locimetric passwords.

### 3.1 Participant Recruitment and Profile

Participants were recruited using Amazon Mechanical Turk, which has become quite popular for usability studies. Our sample was comprised of a total of 204 participants from the United States. The first 102 participants were presented with the high-resolution condition and the subsequent 102 with the low-resolution condition. A range of requirements and level of compensation were used, as attracting participants varied in difficulty during the recruitment process. The breakdown by characteristic and condition (i.e., low- vs. high-resolution) is outlined in Table 1. A series of chi-squared tests indicates that there is no statistically significant difference in the structure of the groups with respect to factor.

Table 1. Participant Profile

| Factor | Category | Low Resolution | High Resolution | Combined |
|---|---|---|---|---|
| Gender | Female | 39 (38.24%) | 41 (40.20%) | 80 (39.22%) |
| | Male | 63 (61.76%) | 61 (59.80%) | 124 (60.78%) |
| Age | 18-30 | 27 (26.47%) | 28 (27.45%) | 55 (26.96%) |
| | 31-45 | 52 (50.98%) | 51 (50.00%) | 103 (50.49%) |
| | 46 or older | 23 (22.55%) | 23 (22.55%) | 46 (22.55%) |
| Education | < Undergraduate degree | 24 (23.53%) | 31 (30.69%) | 55 (27.09%) |
| | Associate's degree | 8 (7.84%) | 6 (5.94%) | 14 (6.90%) |
| | Bachelor's degree | 47 (46.08%) | 50 (49.50%) | 97 (47.78%) |
| | Postgraduate | 23 (22.55%) | 14 (13.86%) | 37 (18.23%) |
| Income | $10,000 - $39,999 | 29 (29.00%) | 34 (33.66%) | 63 (31.34%) |
| | $40,000 - $79,999 | 43 (43.00%) | 45 (44.55%) | 88 (43.78%) |
| | > $80,000 | 28 (28.00%) | 22 (21.78%) | 50 (24.88%) |
| Race | Asian | 11 (10.78%) | 5 (4.90%) | 16 (7.84%) |
| | Black or African American | 6 (5.88%) | 14 (13.73%) | 20 (9.80%) |
| | Other | 0 (0.00%) | 2 (1.96%) | 2 (0.98%) |
| | White | 85 (83.33%) | 81 (79.41%) | 166 (81.37%) |

## 4 Analysis and Results

To identify whether clustering was present, we first generated scatterplots (see Figure 2) to visualize where each of the password points was located on each of the images. This was done using the seaborn visualization package (version 0.11.1) [28]. We then inspected those scatterplots and found clear evidence of clustering. To further support this initial finding, we conducted a series of Clark-Evans tests [9], designed explicitly to identify spatial randomness, using R (version 4.0.3) [1, 2, 24] and, in particular, the spatstat package [3]. One can interpret the results of the Clark-Evans test by examining the R index, as "when R = 0, there is a limit situation of complete aggregation" and then "when R = 1 the pattern of distribution of individuals is random" [15]. To address the edge bias [23] inherent in the original Clark-Evans test, we relied on Donnelly's [12] correction. The results support the conclusion that clustering was present in all of the high-resolution images tested, as the R values were all between 0 and 1 and were statistically significant (see Table 2). Furthermore, the locimetric passwords based on high-resolution images exhibit a slightly lower level of clustering ($\mu = 0.40816$) than the corresponding low-resolution images ($\mu = 0.57608$).

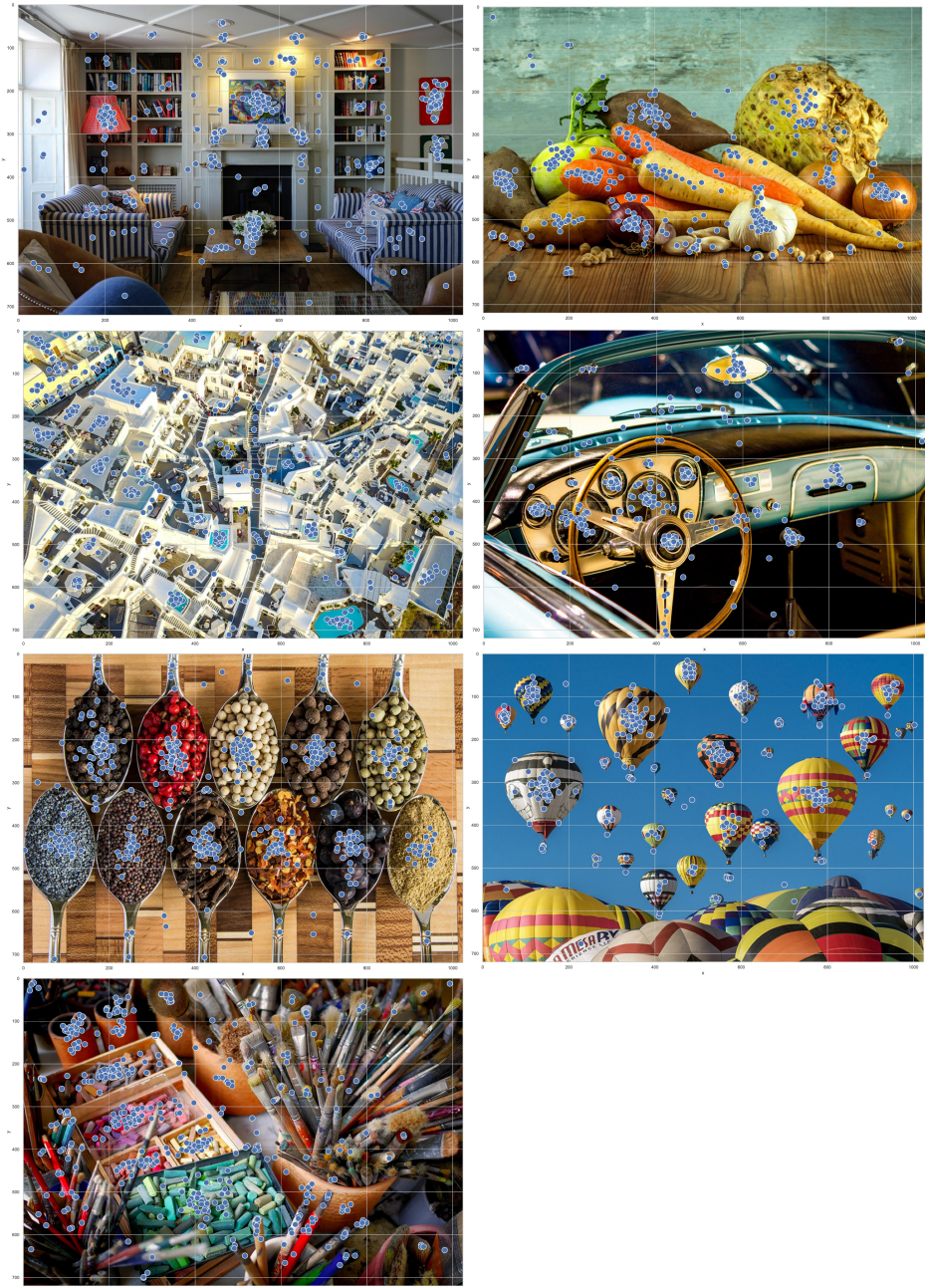

**Fig. 2.** Scatterplot of points that users selected on high-resolution images to form their locimetric password.

Table 2. List of Images and Corresponding Spatial Randomness

| Image | Description | N | R: Low Resolution 5-points | R: High Resolution 3-points | R: High Resolution 5-points |
|---|---|---|---|---|---|
| 1 | Home Interior [18] | 102 | 0.60042 | 0.41253 | 0.40277 |
| 2 | Vegetables [22] | 102 | 0.53065 | 0.38763 | 0.37804 |
| 3 | Landscape [21] | 102 | 0.63515 | 0.40670 | 0.39522 |
| 4 | Vehicle [17] | 102 | 0.52444 | 0.37172 | 0.33962 |
| 5 | Spices [20] | 102 | 0.54077 | 0.51709 | 0.51457 |
| 6 | Hot Air Balloons [19] | 102 | 0.52618 | 0.35804 | 0.33930 |
| 7 | Drawing Tools [16] | 102 | 0.67496 | 0.47499 | 0.48757 |
| Average | | | 0.57608 | 0.41839 | 0.40816 |

Note: For all R values, p < 0.01

## 4.1 Password Length

We also considered the possibility that the use of longer locimetric passwords (i.e., those with more points) might be influencing the presence of hot-spots and resulting in the reuse of the same point(s) multiple times. To make that determination, we compared the presence of clustering within high-resolution images, looking at both the first three points selected by users as part of their passwords and at all five points (see Table 2). We conclude that increasing the points for passwords from three points to five points does not result in a level of clustering that is radically different (i.e., μ = 0.41839 for three-point passwords vs. μ = 0.40816 for five-point passwords). Thus, there is no evidence that longer passwords would lead to the reuse of password points. Conversely, longer locimetric passwords led to slightly less clustering, indicating that the passwords were marginally stronger.

Table 3. Number of Users who Apply Patterns or Reuse the Same Points

| Image | Low Resolution | | | High Resolution | | |
|---|---|---|---|---|---|---|
| | x-dim | y-dim | both dimensions | x-dim | y-dim | both dimensions |
| 1 | 72 | 85 | 8 | 56 | 67 | 4 |
| 2 | 28 | 83 | 5 | 15 | 57 | 3 |
| 3 | 47 | 64 | 8 | 38 | 34 | 5 |
| 4 | 31 | 78 | 8 | 20 | 43 | 8 |
| 5 | 12 | 99 | 4 | 7 | 85 | 5 |
| 6 | 32 | 68 | 5 | 18 | 44 | 4 |
| 7 | 48 | 40 | 4 | 26 | 34 | 2 |
| Average | 38.57143 | 73.85710 | 6.00000 | 25.71430 | 52.00000 | 4.42857 |

A further investigation, this time to identify the number of users who reuse password points within a 10-pixel threshold, revealed that this figure varies by image. For high-resolution images it ranges between 2 and 8 ($\mu = 4.42857$) users. Similarly, for low-resolution images, the number of users was between 4 and 8 ($\mu = 6.00000$) with respect to image. Additionally, we sought to understand whether there were any patterns in the way in which users were selecting their password points. Specifically, we wanted to ascertain whether users would pick points following an imaginary horizontal or vertical line (i.e., within the same column or row of the image), again allowing for a 10-pixel threshold. For high-resolution images, we found that, depending on the image, anywhere from 7 to 85 users would generate passwords in the form of a horizontal line ($\mu = 25.71430$ users) or vertical line ($\mu = 52.00000$ users). For low-resolution images, the use of patterns by users was even more pronounced (this time between 12 and 99 users), with respect to both the horizontal ($\mu = 38.57143$ users) and vertical ($\mu = 73.85710$ users) dimensions. These results are outlined in Table 3.

### 4.2 Considering Demographics

The impact of demographics (i.e., age and gender) on the usability of locimetric passwords was explored for high-resolution five-point passwords. To accomplish this, we once again relied on the Clark-Evans test using Donnelly's edge correction (see Table 4). In the first instance, we compared participants who were over 35 years of age ($\mu = 0.43369$) to those who were 35 or younger ($\mu = 0.42161$). Age was found not to affect randomness, though younger participants generate slightly stronger passwords. In the second instance, we compared the level of clustering between males ($\mu = 0.41029$) and females ($\mu = 0.46251$) but found no consistent pattern amongst all the images evaluated; however, the mean scores indicate that men generate slightly stronger passwords.

**Table 4.** Clustering with Respect to Age and Gender

| Image | R: Age <=35 | R: Age >35 | R: Male | R: Female |
|---|---|---|---|---|
| 1 | 0.41439 | 0.48897 | 0.45468 | 0.45029 |
| 2 | 0.38615 | 0.41509 | 0.33840 | 0.44921 |
| 3 | 0.40558 | 0.37023 | 0.37482 | 0.49303 |
| 4 | 0.34447 | 0.35000 | 0.33546 | 0.44325 |
| 5 | 0.49613 | 0.51432 | 0.52483 | 0.47539 |
| 6 | 0.38198 | 0.38391 | 0.30218 | 0.40336 |
| 7 | 0.52260 | 0.51333 | 0.54163 | 0.52302 |
| Average | 0.42161 | 0.43369 | 0.41029 | 0.46251 |

Note: For all R values, $p < 0.01$

## 5   Discussion, Limitations, and Future Work

Through this work we established that hot-spots persist even with high-resolution images, exploring an extensive collection of images to come to this conclusion. Although the level of clustering was considerable, the findings do indicate that as resolution increases, the level of clustering decreases. The implications of these findings are twofold. In the short term, they indicate that the hot-spot problem should be taken into consideration and addressed by both users and systems administrators so that approaches can be put in place to mitigate the presence of hot-spots. These could be either software-based (as demonstrated by the Cued Click-Points [6] and Persuasive Cued Click-Points [8] solutions) or training-based solutions (which have been demonstrated to be effective in the field of computer security [14, 27, 32]). In the longer term, these findings suggest that, at some point, as screen resolutions increase and locimetric authentication mechanisms adjust to support those higher resolutions, the hot-spot problem may dissipate.

Our research also endeavored to identify whether there were any patterns of user behavior that we could identify as being responsible for the manifestation of hot-spots in the high-resolution images that we were using for the experiment. First, we sought to determine whether user characteristics (i.e., age and gender) influence the formation of the hot-spots that were observed. The benefit of finding such a pattern would inform the allocation of resources (such as training) exclusively to those users. However, no such relationship was found; users generally appear to gravitate towards hot-spots equally. Second, we investigated whether users were reusing points or employing simple (i.e., horizontal or vertical) line patterns as part of their passwords, thereby introducing an element of predictability and further weakening the locimetric scheme. These are practices that we found to vary considerably by image, and slightly by resolution. In particular, point reuse in five-point passwords in the high-resolution condition varied between 1.96% and 7.84%, and pattern use ranged between 6.86% and 83.33%, depending on the image. These practices were more prominent in the low-resolution condition, where point reuse varied between 3.92% and 7.84% and pattern use varied between 11.76% and 97.06%, depending on the image.

However, we would be remiss if we did not point out that further research is necessary to confirm the aforementioned conclusions, specifically the finding that as resolution increases: 1) users rely less on the same points, and consequently hot-spots slowly fade away; 2) point reuse by users falls; 3) password points are less likely to appear in the form of a horizontal or vertical line. Accordingly, we identify two ways in which this line of inquiry could be advanced in the future. The first is to test locimetric authentication with even higher resolutions in order to confirm that the benefits realized from the increase of resolution continue and do not plateau. The second involves carrying out a field study that examines usage 'in the wild'. It is also evident that the composition of the image influences the level of clustering; this has been previously highlighted as a consideration regarding the suitability of images for use with locimetric authentication [11]. In our work, we controlled for complexity by taking considerable efforts to ensure that all of our images had a plethora of points for users to select. Nev-ertheless, differences were observed in the level of clustering between images. Consequently, a superior understanding of the interplay between composition and resolution is needed to understand better the benefits that can be realized by the increasing of image resolution.

Should the hot-spot problem be successfully addressed, the locimetric scheme has considerable potential to serve as a competitive alternative to traditional forms of authentication.

**References**

1. Baddeley, A. et al.: Hybrids of Gibbs point process models and their implementation. Journal of Statistical Software. 55, 11, 1–43 (2013). https://doi.org/10.18637/jss.v055.i11.
2. Baddeley, A. et al.: Spatial Point Patterns: Methodology and Applications with R. Chapman and Hall / CRC Press, London (2015).
3. Baddeley, A., Turner, R.: Spatstat: An R package for analyzing spatial point patterns. Journal of Statistical Software. 12, 6, 1–42 (2005).
4. Blonder, G.E.: Graphical password. Patent number: 5559961. United States Patent and Trademark Office (1996).
5. Bulling, A. et al.: Increasing the security of gaze-based cued-recall graphical passwords Using saliency masks. In: Proceedings of the SIGCHI Conference on Human Factors in Computing Systems. pp. 3011–3020 ACM Inc, New York, NY, USA (2012). https://doi.org/10.1145/2207676.2208712.
6. Chiasson, S. et al.: Graphical password authentication using Cued Click Points. In: Biskup, J. and López, J. (eds.) Computer Security – ESORICS 2007. pp. 359–374 Springer Berlin Heidelberg, Berlin, Heidelberg (2007).
7. Chiasson, S. et al.: Influencing users towards better passwords: Persuasive Cued Click-Points. In: Proceedings of the 22nd British HCI Group Annual Conference on People and Computers: Culture, Creativity, Interaction - Volume 1. pp. 121–130 BCS Learning & Development Ltd., Swindon, GBR (2008).
8. Chiasson, S. et al.: Persuasive Cued Click-Points: Design, implementation, and evaluation of a knowledge-based authentication mechanism. IEEE Transactions on Dependable and Secure Computing. 9, 2, 222–235 (2012). https://doi.org/10.1109/TDSC.2011.55.
9. Clark, P.J., Evans, F.C.: Distance to nearest neighbor as a measure of spatial relationships in populations. ecology. 35, 4, 445–453 (1954). https://doi.org/10.2307/1931034.
10. De Angeli, A. et al.: Is a picture really worth a thousand words? Exploring the feasibility of graphical authentication systems. International Journal of Human-Computer Studies. 63, 1–2, 128–152 (2005). https://doi.org/10.1016/j.ijhcs.2005.04.020.
11. Dirik, A.E. et al.: Modeling user choice in the PassPoints graphical password scheme. In: Proceedings of the 3rd Symposium on Usable Privacy and Security. pp. 20–28 ACM Inc., Pittsburg, PA (2007). https://doi.org/10.1145/1280680.1280684.
12. Donnelly, K.: Simulation to determine the variance and edge-effect of total nearest neighbour distance. In: Hodder, I. (ed.) Simulation Methods in Archeology. pp. 91–95 Cambridge University Press, Cambridge, UK (1978).
13. Gao, H. et al.: The hot-spots problem in Windows 8 graphical password scheme. In: Wang, G. et al. (eds.) Cyberspace Safety and Security. pp. 349–362 Springer International Publishing (2013).
14. Huang, D.-L. et al.: A survey of factors influencing people's perception of information security. In: Proceedings of the 12th International Conference on Human-computer Interaction: Applications and Services. pp. 906–915 Springer-Verlag, Beijing, China (2007).
15. Petrere, M.: The variance of the index (R) of aggregation of Clark and Evans. Oecologia. 68, 1, 158–159 (1985). https://doi.org/10.1007/BF00379489.

16. Pixabay: Brush Chalk Color Atelier Paint. https://pixabay.com/photos/brush-chalk-color-atelier-paint-2927793/. Accessed 2 February 2021.
17. Pixabay: Car Vehicle Motor Transport. https://pixabay.com/photos/car-vehicle-motor-transport-3046424/. Accessed 2 February 2021.
18. Pixabay: Home Interior Room House Furniture. https://pixabay.com/photos/home-interior-room-house-furniture-1438305/. Accessed 2 February 2021.
19. Pixabay: Hot Air Balloons Adventure Balloons. https://pixabay.com/photos/hot-air-balloons-adventure-balloons-1867279/. Accessed 2 February 2021.
20. Pixabay: Mat Spices. https://pixabay.com/photos/mat-spices-3251064/. Accessed 2 February 2021.
21. Pixabay: Santorini City Greece Tourism. https://pixabay.com/photos/santorini-city-greece-tourism-4044972/. Accessed 2 February 2021.
22. Pixabay: Vegetables Carrots Garlic Celery. https://pixabay.com/photos/vegetables-carrots-garlic-celery-1212845/. Accessed 2 February 2021.
23. Pommerening, A., Stoyan, D.: Edge-correction needs in estimating indices of spatial forest structure. Canadian Journal of Forest Research. 36, 7, 1723–1739 (2006). https://doi.org/10.1139/x06-060.
24. R Core Team: A language and environment for statistical computing. R Foundation for Statistical Computing, Vienna, Austria (2013). http://www.R-project.org/.
25. Stobert, E. et al.: Exploring usability effects of increasing security in click-based graphical passwords. In: Proceedings of the 26th Annual Computer Security Applications Conference. pp. 79–88 ACM Inc., New York, NY (2010). https://doi.org/10.1145/1920261.1920273.
26. Thorpe, J., Van Oorschot, P.C.: Human-seeded attacks and exploiting hot-spots in graphical passwords. In: Proceedings of the 16th USENIX Security Symposium. pp. 103–118 (2007). https://www.usenix.org/legacy/events/sec07/tech/full_papers/thorpe/thorpe.pdf. Accessed 11 June 2021.
27. Ur, B. et al.: How does your password measure up? The effect of strength meters on password creation. In: Proceedings of the 21st USENIX Security Symposium. pp. 65–80 (2012). https://www.usenix.org/system/files/conference/usenixsecurity12/sec12-final209.pdf. Accessed 11 June 2021.
28. Waskom, M. et al: mwaskom/seaborn: v0.11.1. Zenodo (2020). https://zenodo.org/record/4379347#.YMQHVjZKh6M. Accessed 11 June 2021.
29. Wiedenbeck, S. et al.: Authentication using graphical passwords: Basic results. In: Proceedings of the 11th International Conference on Human-Computer Interaction International (HCII 2005). Las Vegas, NV (2005). http://www.jimwaters.info/pubs/Graphical-Password-Basic-Results-2005.pdf. Accessed 11 June 2021.
30. Wiedenbeck, S. et al.: Authentication using graphical passwords: Effects of tolerance and image choice. In Proceedings of the 2005 Symposium on Usable Privacy and Security (2005). pp. 1-12 (2005). https://doi.org/10.1145/1073001.1073002.
31. Wiedenbeck, S. et al.: PassPoints: Design and longitudinal evaluation of a graphical password system. International Journal of Human-Computer Studies. 63, 1, 102–127 (2005). https://doi.org/10.1016/j.ijhcs.2005.04.010.
32. Yıldırım, M., Mackie, I.: Encouraging users to improve password security and memorability. International Journal of Information Security. 18, 6, 741–759 (2019). https://doi.org/10.1007/s10207-019-00429-y.